\documentclass[symmetry,article,accept,oneauthor]{Definitions/mdpi} 

\pdfoutput=1

\firstpage{1} 
\pubvolume{12}
\issuenum{4}
\articlenumber{597}
\pubyear{2020}
\copyrightyear{2020}
\history{Received: 8 March 2020; Accepted: 24 March 2020; Published: 9 April 2020.}

\def\a{\alpha}
\def\b{\beta}

\def\e{\varepsilon}

\def\l{\lambda}

\def\ra{\rightarrow}

\usepackage{amssymb}
\usepackage{bm}

\Title{Scale Symmetry in the Universe}

\Author{{Jose Gaite} 
\orcidA{}}

\AuthorNames{Jose Gaite}

\address[1]{%
Applied Physics Dept.,
ETSIAE, Univ.\ Polit\'ecnica de Madrid, E-28040 Madrid, Spain; jose.gaite@upm.es\\[3mm]
In memory of my mother.}

\corres{Correspondence: jose.gaite@upm.es; Tel.: +x-xxx-xxx-xxxx}

\abstract{Scale symmetry is a fundamental symmetry of physics that seems however 
not to be fully realized in the universe. Here,   we focus on the astronomical 
scales ruled by gravity, where scale symmetry holds and gives rise to a truly 
scale invariant distribution of matter, namely  it gives rise to a fractal geometry. 
A suitable explanation of the features of the fractal cosmic mass distribution 
is provided by the nonlinear Poisson--Boltzmann--Emden equation. 
An alternative interpretation of this equation is connected with theories of quantum gravity.
We study the fractal solutions of the equation and connect them with 
the statistical theory of random multiplicative cascades, which originated in the theory 
of fluid turbulence. 
The type of multifractal mass distributions so obtained agrees with  
results from the analysis of cosmological simulations and of observations of the galaxy distribution.}


\begin{document}

\section{Introduction}

The symmetry of the physical laws is probably the essential foundation of our 
current understanding of physics and the universe~\cite{Feynman}.
Symmetry principles are indeed essential in the formulation of quantum field theory, 
as one of the fundamental theories of physics~\cite{Weinberg}.
The oldest and most common symmetries are the space-time symmetries, namely   the 
symmetry of the physical laws under space or time translations and under space rotations, 
a symmetry that is enlarged to the Poincar\'e symmetry group by the theory of relativity. 
These symmetries induce very relevant conservation laws, namely   the conservation of linear
and angular momenta and the conservation of energy. \mbox{When a theory} of gravity is added to quantum field theory, 
the space-time symmetries become more involved, because~they only hold locally in inertial frames. 
However, the~relativistic 
theories of gravity can be formulated as {\em gauge} theories of 
the space-time symmetries~\cite{Hehl}. In~addition to the mentioned space-time symmetries, 
there is another transformation of space-time
that is intuitively appealing and has had an important role in physics and other sciences, 
even though it is not necessarily a symmetry, namely   the transformation of scale or dilatation. It closes, together 
with the space-time transformations, a~group that can be enlarged to 
the {\em conformal group} of transformations, 
of ever increasing interest in theoretical physics~\cite{Kastrup}.

Given the large range of sizes in the universe~\cite{10}, it is surely an old question 
how to know what determines the relevant sizes, from~micro to macro-physics. While 
micro-physics or human scale physics involve various physical constants and laws, 
macro-physics, especially  the large scale structure of the universe, is just the realm 
of gravity. Gravity has no intrinsic length scale, so one can wonder why large astronomical 
objects have a given size. In~fact, beyond~galaxies, whose size is determined by both 
gravity and the electromagnetic interaction~\cite{Padma}, there seems to be no way 
to construct objects of \mbox{a given} size. Simply put, if~one finds an object of a given 
size, there must be similar objects of larger size (and possibly of smaller size, as~long as 
one does not go to too small scales). For~example, take a cluster of galaxies; there 
must be similar {\em superclusters} of every possible size. 
Not surprisingly, the~idea of \mbox{a scale} invariant structure of the universe on large scales 
is old, but~its modern formulation had to await the advent of the appropriate 
mathematical description, namely   fractal geometry~\cite{Mandel}. Simple fractals are 
scale invariant and are indeed composed of clusters of clusters of \dots, down to 
the infinitesimally small. Naturally, in~the universe, the~self-similarity must stop 
at a scale about the size of galaxies,  
although it could be limitless towards the large scales, in~principle. 

However, the~appealing idea of an infinite hierarchy of clusters of clusters 
of galaxies~\cite{Vaucou,Mandel} clashes with the large scale homogeneity 
prescribed by the standard cosmological principle and embodied in 
the Friedmann--Lemaitre--Robertson--Walker relativistic model of the universe~\cite{Pee,Peebles}.
\mbox{Naturally, a~compromise} is possible: the universe is homogeneous on very large 
scales but is fractal on smaller yet large scales 
(in the so-called strong-clustering regime). In~the intermediate range of scales, 
the structure of matter in the universe undergoes a transition from fractal to homogeneous. 
Therefore, there is \mbox{a scale} of transition to homogeneity, which admits several definitions 
that should give approximately the same value. However, 
despite   the work of many researchers along several decades, 
the debate about the scale of transition to homogeneity 
is not fully settled and quite different values appear in the literature~\cite{Pee,Peebles,Cole-Pietro,Borga,Sylos-PR,Jones-RMP,I0}. 
This situation is surely a consequence of the different definitions used and 
the limitations of the current methods of observation. At~any rate, this mainly 
observational issue is not crucial for us and we are content to study the fractal 
structure of the universe without worrying about the definitive value of 
the scale of transition to~homogeneity.

It is normal in physics that scale invariance holds in one range of scales and 
is lost in another range or changes to a different type of scale invariance 
(in a sense, the~homogeneous state is trivially scale invariant). 
This situation is common, for~example, 
in critical phenomena in statistical physics, in~which it is called {\em crossover}. 
The theory of critical phenomena is actually a fruitful domain of application of 
the theory of scale invariance and furthermore of the full theory of conformal
invariance~\cite{Henkel}. The~purpose of the present paper is to study the theory 
of the fractal structure of the universe with methods of statistical physics 
and field theory. There is a basic difference between the large mass fluctuations in 
fractal geometry and the more moderate fluctuations in the theory of critical phenomena~\cite{I0}, but~statistical field theory methods can nonetheless be applied to 
fractal geometry. At~any rate, the~scale invariance and fractal nature of 
gravitational clustering is due to the 
form of the law of
gravity. In~fact, the~statistical field theory of gravitational systems is~peculiar.

The Newton law, with~its inverse dependence on distance, does not fulfill the 
condition of short-rangedness that makes an interaction
tractable in statistical physics,  thus there can be no homogeneous equilibrium state
(\S 74, \cite{LL}). 
This problem leads to the peculiar statistical and thermodynamic properties of 
many-body systems with long-range interactions, such as negative specific heats, 
ensemble nonequivalence, metastable states whose lifetimes diverge with the number of bodies, 
and spatial inhomogeneity, among~others~\cite{Chava}. Fortunately, there are methods 
to study gravitationally interacting systems. 
In the mean field approximation for a system of $N$ bodies in gravitational interaction, which becomes exact for $N \ra \infty$, the~equation that rules the distribution of the 
gravitational potential is a higher dimensional generalization of Liouville's equation, 
\mbox{originally introduced} to describe the conformal geometry of surfaces~\cite{Liouville}. 
The gravitational equation is also old and has been given various names;  
it is called the Poisson--Boltzmann--Emden equation in Bavaud's review~\cite{Bavaud}. 
Its scale covariance is remarked by de Vega~et~al.~\cite{deVega}
(who actually studied  the fractal structure of the interstellar medium, on~galactic scales 
below the cosmological scales).

Deep studies of the fractal geometry of mass distributions related to the 
Poisson--Boltzmann--Emden equation have been carried out in the theory of 
stochastic processes, namely   in the theory of 
random multiplicative cascades~\cite{MFcascades,LiouvilleQG,Mchaos}. 
Historically, this theory arose in relation to the {\em lognormal model} 
of turbulence~\cite{Kolmo62,Frisch}. 
Random multiplicative cascades give rise to multifractal distributions, which 
have applications in several areas~\cite{MandelMF,Harte}. 
Indeed, random multiplicative cascades are naturally applied to cosmology: while 
in models of turbulence the energy or vorticity ``cascades'' down towards 
smaller scales, in~cosmology, mass undergoes successive gravitational collapses. 
The result of the infinite iteration of random mass condensations 
is \mbox{a multifractal} mass distribution, that is to say, a~generalization 
of a simple fractal~structure.   

Gravity is the dominant force in the universe on large scales but it also dominates 
the very small scales near the Planck length, which is the domain of quantum gravity~\cite{QG}.
Field equations connected with the Poisson--Boltzmann--Emden equation appear in 
generalizations of the general theory of relativity that add a scalar field, such as 
the Dicke--Brans--Jordan theory~\cite{QG}. Scalar--tensor theories of gravity are old but 
they gained popularity with the advent of string theory, a~theory of 
quantum gravity in which a scalar field, the~dilaton, naturally arises as a partner of 
the graviton (the metric tensor field quantum) \cite{Polyakov,GSW}. The~dilaton is essential 
in the understanding of the conformal symmetry of space-time~\cite{tHooft}. However, 
the connection of the dilaton field with the gravitational potential in the 
Poisson--Boltzmann--Emden equation is, at~best, indirect; except in two-dimensional 
relativistic gravity, which is defined in terms of only one scalar field~\cite{Polyakov}. Of~course, the~fractal geometry of the 
large scale structure of the universe is produced by three-dimensional gravity, but~
the study of lower dimensional models also has interest in~cosmology.

We begin with a summary of fractal geometry, in~particular, of~
multifractal geometry, \mbox{orientated towards} the description of the cosmic mass distribution. 
We proceed to study how to generate the fractal geometry of the 
large scale structure, especially  combining the Vlasov dynamics and the 
Poisson--Boltzmann--Emden equation~\cite{Bavaud} with 
the Zel'dovich approximation and the adhesion model of the early stage 
of structure formation~\cite{Shan-Zel,GSS}. 
The fractal geometry of the web structure that arises in the cosmological evolution 
according to the adhesion model is thoroughly studied in~\cite{AinA}. 
\mbox{In this} paper, we take a further step in the attempt to describe analytically 
the fractal geometry of the universe, by~means of simple models and taking advantage 
of the power of the scale~symmetry. 

\section{Multifractal~Geometry}
\label{MFgeom}

Multifractal geometry is, of~course, the~geometry of multifractal mass distributions, which 
are a natural generalization of the concept of simple fractal set, such that 
a range of dimensions appear instead of just a single dimension. 
However, multifractal geometry can almost be defined as the geometry of 
{\em generic} mass distributions, 
because most of them are actually susceptible to a multifractal analysis. To~be precise, 
despite common prejudices, 
generic mass distributions are {\em strictly singular}, \mbox{that is} to say, the~mass density 
is not well defined and is in fact either zero or infinity at every point~\cite{Monti}.  
The singularities give rise to a spectrum of dimensions, as~indeed happens in 
the mass distributions that appear in cosmology~\cite{AinA}. 

Actually, the~formation of structure in the universe occurs in a definite way, from~
the growth of small density fluctuations in an initially homogeneous and 
isotropic universe, up~to a size such that the increased gravitational force leads 
to a collapse of mass patches towards the formation of singularities. 
In the adhesion model of structure formation, 
the collapse initially leads to matter sheets (two-dimensional singularities), then to 
filaments (one-dimensional singularities), and~finally to point-like singularities
~\cite{Shan-Zel,GSS}. The~web structure so formed is a particular example 
of multifractal mass distribution and is indeed a rough model of the actual 
cosmic structure, but~it is not a very accurate model~\cite{AinA}. The~problem is that 
the formation of matter filaments and especially point-like singularities, in~Newtonian 
gravitation, can only occur after the dissipation of an 
infinite amount of gravitational energy, so that the process requires 
a fully relativistic treatment (Box 32.3, \cite{QG}), beyond~the scope of the adhesion~model.

\begin{figure}
\centering
\includegraphics[height=4.95cm]{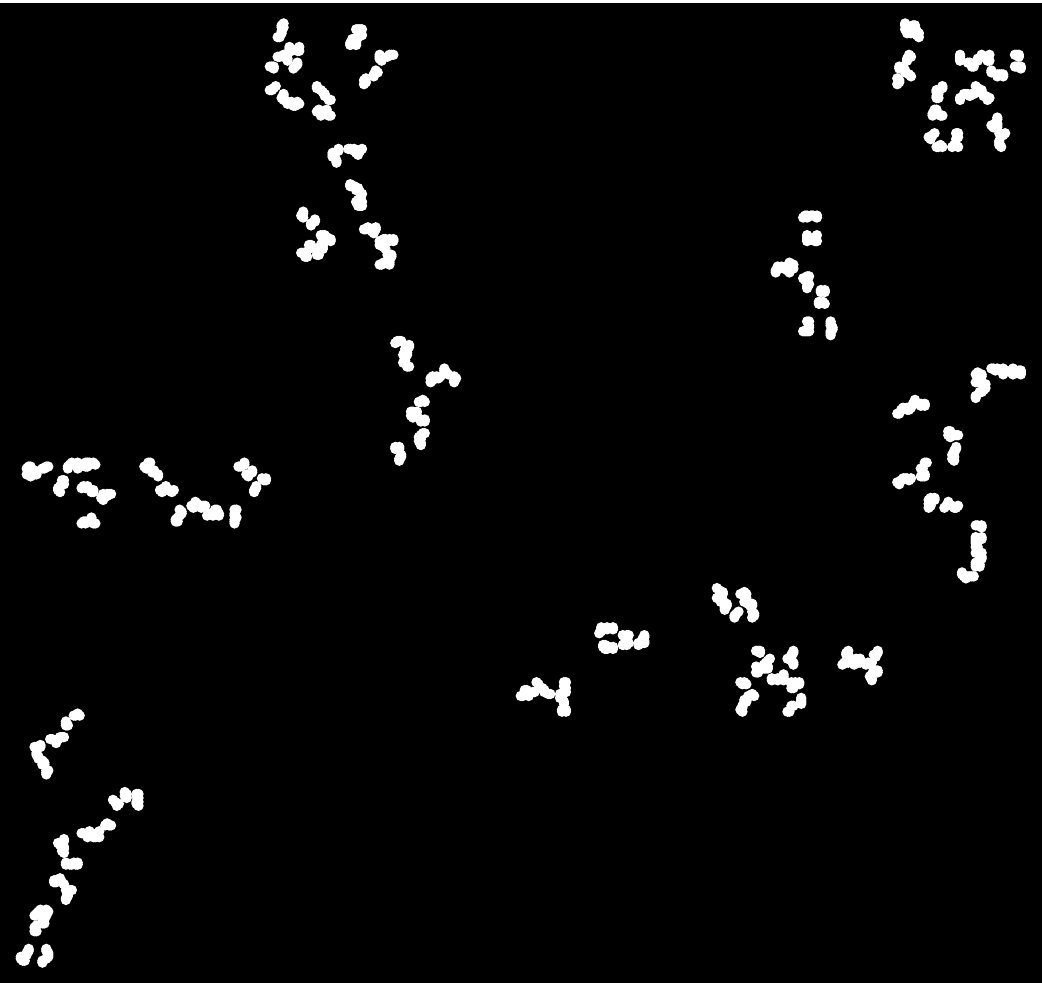}
\includegraphics[height=4.95cm]{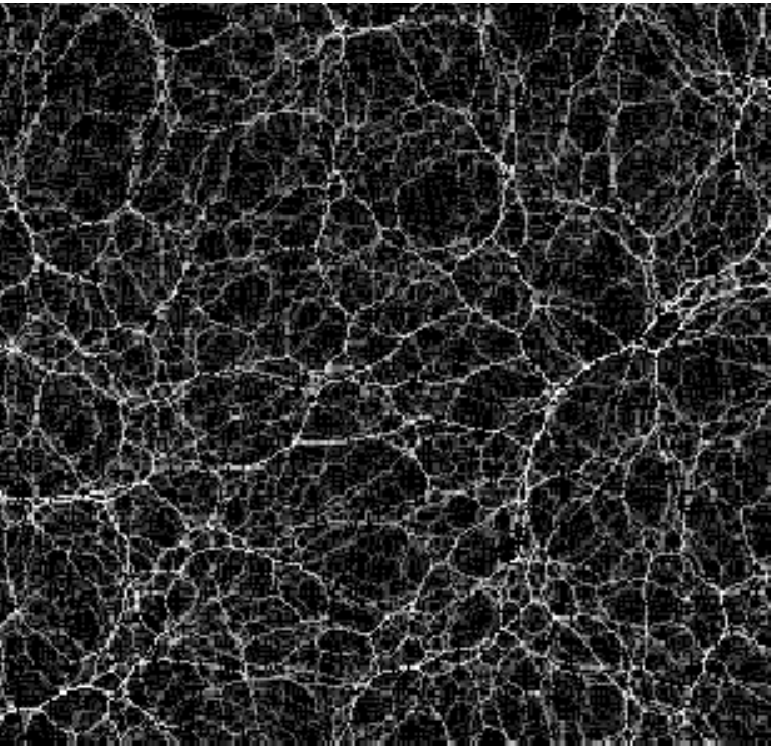}
\includegraphics[height=4.95cm]{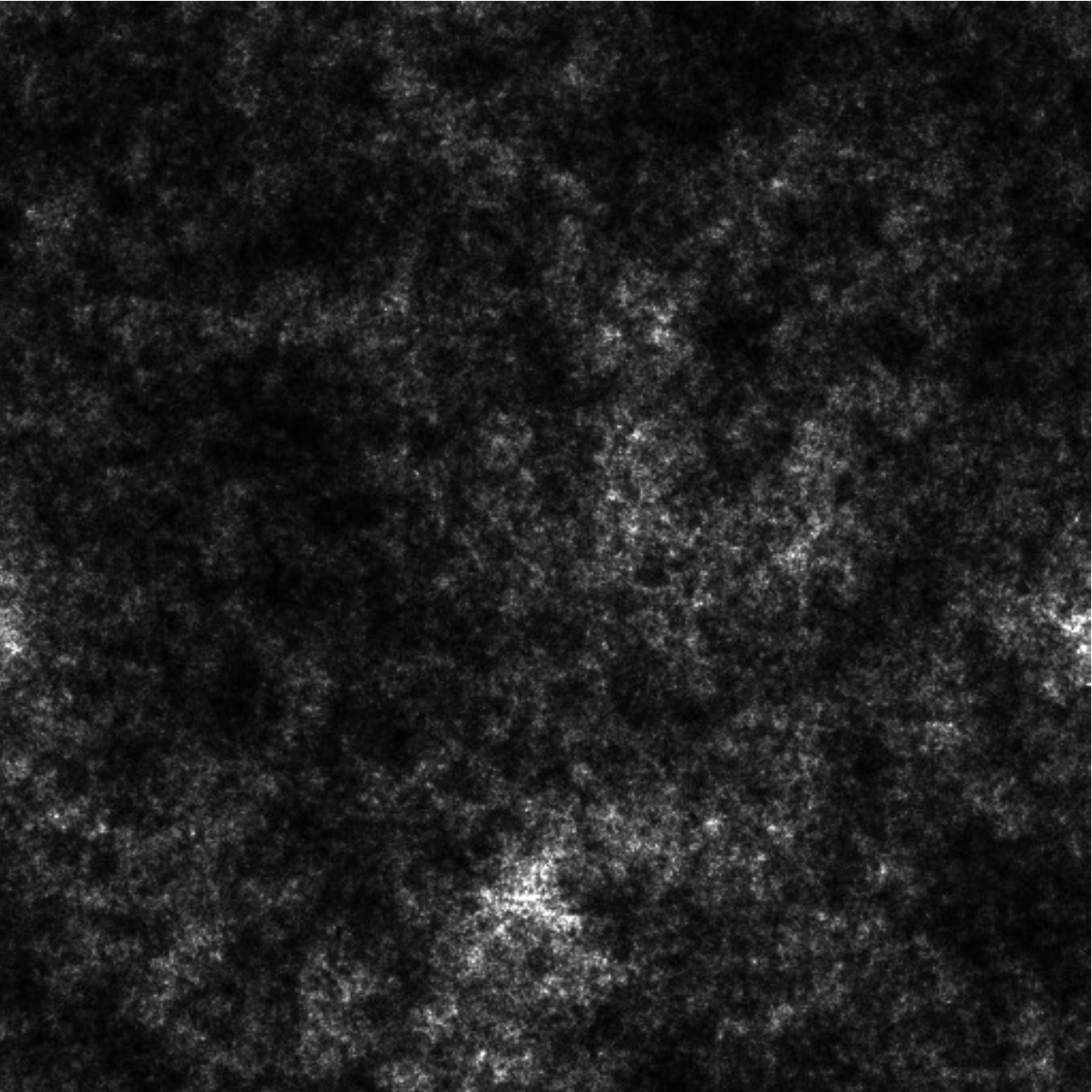}
\caption{Three different types of fractal structure that can represent the cosmic 
mass distribution (from left to right): (i) monofractal cluster hierarchy, with~$D=1$ and 
therefore $m(r) \propto r$;  (ii) cosmic web structure, rendered in two dimensions, showing 
one-dimensional filaments and zero-dimensional nodes organized in a 
self-similar structure; and  (iii) 
multifractal lognormal mass distribution, \mbox{with~a range} of dimensions $\a$ such that 
$m(r,\bm{x}) \propto r^{\a(\bm{x})}$ from the point $\bm{x}$. 
}
\label{devil}
\end{figure}

At any rate, the~adhesion model is not meant to describe accurately the formation 
of singularities, even within Newtonian gravitation. 
Gurevich and Zybin's specific approach to this process 
(within Newtonian gravitation) \cite{GZ} obtains 
different results, namely   the formation of singular power-law mass concentrations 
instead of the collapse of a number of dimensions (one, two or three) to zero size.
Singular power-law mass concentrations are the hallmark of typical multifractals. 
On the other hand, multifractals are the natural generalization of simple hierarchical mass 
distributions, namely   simple fractal sets, also referred to as  {\em monofractal} distributions.
Therefore, we are led to study the geometry of multifractal mass distributions and 
the proper way to characterize them~\cite{AinA}. Examples of monofractal, multifractal 
and adhesion model web structures are shown in Figure~\ref{devil}.

The mathematical definition of multifractal can be found in books about 
fractal geometry~\cite{Harte,Falcon}. It is to be remarked that the definition of 
multifractal, as~well as the simpler definition of fractal set, are formulated 
in terms of limits for some vanishing length scale. In~other words, 
the scale symmetry only needs to hold asymptotically for vanishing length scales. However, 
there is an important class of multifractals in which scale symmetry holds 
in a finite range of scales, namely   the {\em self-similar} fractals or multifractals, 
generated by some iterative process~\cite{Harte,Falcon}. These processes 
are especially interesting in the theory of cosmological structure formation. 
One reason for it is the symmetry of the initial state, which is strictly homogeneous and 
isotropic and is such that structure formation proceeds homogeneously in every part 
of the universe, defining a single scale of transition to homogeneity. 
Another reason, related to the preceding one, is that the standard 
cosmological principle admits \mbox{a weaker}, stochastic formulation, namely   
Mandelbrot's conditional cosmological principle, in~which every {\em possible} observer 
is replaced by every observer located at a {\em material} point 
(\S 22, \cite{Mandel}). 
This principle implies the self-similarity of the 
strong-clustering~regime.

As mentioned  in the  Introduction, 
good models of the strong-clustering regime of the cosmological evolution, 
in terms of successive gravitational collapses, 
are the random multiplicative cascades. 
\mbox{Before we} study in detail these cascade processes in Section~\ref{cascades}, we prepare 
ourselves by studying 
methods of multifractal analysis, especially those that take into account 
the statistical homogeneity and isotropy of the mass~distribution. 

A statistically homogeneous, isotropic, and~scale invariant mass distribution can be 
characterized 
by the probability of a density field $\rho(\bm{x})$ with those symmetries, that is to say, 
by a  suitable functional of the function $\rho(\bm{x})$. 
If we forgo the scale invariance, the~simplest functional is, 
of course, the~Gaussian functional, which in Fourier space is simply a product of 
the respective Gaussian probability functions over all the Fourier modes~\cite{Padma}. 
However, in~cosmology, it is normal to use, instead of the probability of $\rho(\bm{x})$, 
the $N$-point correlation functions of $\rho(\bm{x})$ 
or the method of {\em counts in cells} \cite{Peebles}. 
These counts are usually galaxy counts in a sphere of radius $r$ placed at random
across a fair sample of the process~\cite{Peebles}. 
Therefore, the~density refers to just the galaxy number density. 
This method of galaxy counts in cells has been applied to 
the study of scale invariance in the large scale structure of the universe~\cite{Bal-Scha}. 
\mbox{However, galaxy counts} do not consider the mass of galaxies. Nevertheless, 
in cosmological $N$-body simulations, which describe the mass distribution 
in terms of $N$ bodies (or particles) of equal mass, number density equals mass density, 
so the density is the real mass density or, to~be precise, a~{\em coarse-grained} mass density. Furthermore, the~good mass resolution of the simulations 
makes them suitable for the multifractal analysis~\cite{Valda,Colom,Yepes,fhalos,I4}. 
The results of these analyses are summarized in Section~\ref{data}.


It is often assumed that the full set of $N$-point correlation functions allow   recovering  
the probability functional of $\rho(\bm{x})$ or that 
the full set of integral statistical moments of the 
probability distribution function of the coarse-grained density allow   recovering  
the probability distribution function of the coarse-grained density. Both assumptions 
do not necessarily hold for random multifractals. 
\mbox{Surely, this is a problem} 
that affects the strictly singular distributions, as~they have  
mass densities that are zero or infinity at every point. It is clear that this 
singular behavior makes the probability distribution function of 
the coarse-grained density singular for a vanishing {\em coarse-graining length} (e.g., 
for a vanishing radius $r$ of the above-mentioned sphere): 
the probability distribution function gets 
concentrated on zero or infinity. In~fact, the~multifractal analysis can be made 
in terms of the probability distribution function of the coarse-grained density, 
analyzing {\em how} it becomes singular. This type of multifractal analysis is 
equivalent to the lattice or point centered based methods~\cite{Harte,Falcon}, 
according to the type of coarse-graining, if~we take into account the homogeneity 
and isotropy of the mass distribution. Of~course, 
the mass distribution must also be self-similar, namely   
scale invariant in some range of scales and not just asymptotically for 
vanishing coarse-graining~length. 

It has to be remarked here that self-similarity is not equivalent to multifractality. 
A pertinent example is the case of the adhesion model with scale invariant 
initial conditions~\cite{V-Frisch}. The~scale invariant initial conditions consist of 
a uniform density and a {\em fractional Brownian} velocity field. \mbox{This type} 
of field is self similar but not multifractal; in fact, it is a normal non-differentiable 
function (ch.\ IX, \cite{Mandel}) (\mbox{ch.\ 16, \cite{Falcon}}). However, in~the adhesion model, 
the mass distribution becomes singular and multifractal as \mbox{a consequence} of the 
scale invariant initial conditions~\cite{V-Frisch,AinA}.


In a multifractal distribution, the~point density $\rho(\bm{x})$ is singular: in 
mathematical terms, it does not exist as a function.  
However, the~coarse-grained density is a standard function, namely   the coarse graining 
is a {\em regularization} of $\rho(\bm{x})$. However, it is a function that depends 
on the coarse-graining length, say $r$, and~on the precise method of coarse graining. 
Since the point density $\rho(\bm{x})$ is a random variable,  thus is the coarse-grained density. 
The probability distribution of this variable is independent of the base point, by~
statistical homogeneity, and~it is a suitable starting point for the multifractal analysis, 
\mbox{as said} above. 
Let us call $\rho_r$ the density coarse grained with length $r$ and 
$P(\rho_r)$ its probability distribution function. We are interested in the 
statistical moments of this probability distribution function. \mbox{These moments} are 
related to the density correlation functions: in fact, the~statistical moment of 
order $k$ can be expressed as the integral of the $k$-point correlation function over the 
$k$ points in the coarse-graining volume. Furthermore, there is a relation between 
the probability of given counts in cells in some volume and the statistical moments 
of the probability distribution function of the number density in that volume~\cite{Bal-Scha,White}.
In particular, the~probability of the volume being empty, called the {\em void probability 
function}, is the Laplace transform of the probability distribution function of 
the density and can be taken as the generating function of the statistical moments 
of this probability distribution function, following standard methods of statistical 
mechanics. Naturally, these relations play a role in the study of {\em cosmic voids}
\cite{voids}. Here, we focus on just the probability $P(\rho_r)$.

The multifractal analysis consists in finding how $P(\rho_r)$ depends on $r$ 
through its {\em fractional} statistical moments. Within~the {\em scaling regime},  
every fractional statistical moment behaves as a power law of $r$, namely   
for $q \in \mathbb{R}$,
\begin{equation}
\mu_q = \frac{\langle {\rho_r}^q\rangle}{\langle {\rho_r}\rangle^q} = 
\int_0^\infty \left(\frac{{\rho_r}}{\langle {\rho_r}\rangle}\right)^q P(\rho_r) d\rho_r 
\propto r^{\tau(q)-3(q-1)}, 
\label{tq}
\end{equation}
where the somewhat complicated exponent of $r$ comes from the somewhat different definition 
of statistical moments that is standard in fractal geometry~\cite{I4,AinA}. 
Notice that we assume that the multifractal has support in the full three-dimensional space, 
that is to say, we have a {\em nonlacunar} multifractal, with~$\tau(0)=-3$, 
as is natural in Newtonian cosmology. 
The function $\tau(q)$ describes the singularity structure of 
any realization of $\rho(\bm{x})$, that is to say, the~local behavior of the 
mass distribution at every singularity. When this description is complete  
the distribution is said to fulfill the multifractal formalism~\cite{Harte}. 
The local behavior of a realization of the mass distribution at a singular point 
$\bm{x}$ is given by the {\em local dimension} $\a(\bm{x})$, 
which defines how the mass grows from that point, that is to say,
\begin{equation}
m(\bm{x},r) \sim r^{\a(\bm{x})}. 
\label{mr}
\end{equation}

Naturally, $\a \geq 0$. Every set of points with a given local dimension $\a$ constitutes 
a fractal set with a dimension that depends on $\a$, namely   $f(\a)$.
The multifractal formalism involves a relationship between $\tau(q)$
and the local behaviors given by $\a$ and $f(\a)$, in~the form of a Legendre
transform:
\begin{equation}
f(\a) = q\a - \tau(q), 
\label{fa}
\end{equation}
and $\a(q) = \tau'(q)$. 
It is to be remarked that the multifractal formalism becomes trivial 
if there is only one dimension $\a=f(\a)$ 
(the case of a {\em monofractal} or {\em unifractal}); then, $\a=\tau(q)/(q-1)$, 
as follows from Equation~(\ref{fa}).  
The quotient $D_q=\tau(q)/(q-1)$ is called, in~general, the~R\'enyi dimension and has 
an information-theoretic meaning~\cite{Harte}. In~a monofractal, $D_q=\a$ is constant, but~
in a multifractal is \mbox{a non-increasing} function of $q$. 
A typical example of monofractal is a self-similar fractal {\em set} such 
that the mass is uniformly spread on~it.

Before studying the theory of random multiplicative cascades, let us consider 
some simple examples of self-similar multifractals~\cite{Mandel,Harte,Falcon}. They are 
constructed as self-similar fractal sets with \mbox{an irregular} mass distribution on them. 
The simplest example is surely the ``Cantor measure'' (\mbox{\S 1.2.1, \cite{Harte}})
(example 17.1, \cite{Falcon}). If~the mass is distributed on the three subintervals 
instead of just two of them, we have the ``Besicovitch weighted curdling'', 
an example of {\em nonlacunar fractal} (p.~377, \cite{Mandel}).
\mbox{A simpler} nonlacunar fractal is obtained by using just two equal subintervals, 
so defining the {\em binomial multiplicative process} (\S 6.2, \cite{Feder}). 
Its function $\tau(q)$ is very simple:
\begin{equation}
\tau(q) = -\log_2 [p^q+(1-p)^q],
\label{tqm}
\end{equation}
where $p$ is, say, the~mass fraction in the left hand side subinterval.
The generalization of these constructions leads 
to the general theory of deterministic self-similar multifractals (\S 17.3, \cite{Falcon})
(or Moran cascade processes (\S 6.2, \cite{Harte})). 
Notice that the average in Equation~(\ref{tq}), in~the case of deterministic multifractals, 
is to be interpreted as a spatial average in a suitable domain.
\mbox{Random multiplicative} cascades are just an elaboration of the deterministic 
cascades, 
in the setting of random processes. Only in this setting, we can achieve 
statistically homogeneous, isotropic and scale invariant mass~distributions.

Several properties of the self-similar multifractals defined by 
an iterated function system and a set of weights for the redistribution of mass 
are relatively easy to prove (these multifractals are sometimes called 
{\em multinomial} because one iteration gives rise to a multinomial distribution) \cite{Harte,Falcon}. In~a multifractal with support in $\mathbb{R}^3$, $\tau(q)$ is a concave 
and increasing function, such that $\tau(0)=-3$ and $\tau(1)=0$.
Furthermore, $\tau(q)$ has definite asymptotes as $q\ra \infty$ or $q\ra -\infty$ 
(see the realistic example of Figure~\ref{Bolshoi}).
The slopes can be computed and, in~particular,
\begin{equation}
\lim_{q\ra \infty}\frac{\tau(q)}{q} = \a_\mathrm{min},
\end{equation}
where $\a_\mathrm{min}$ is, of~course, the~minimum value of the local dimension and is 
smaller than 3 (or smaller than the ambient space dimension, in~general). 
Therefore, the~asymptotic value of the exponent of $r$ in Equation~(\ref{tq}) for $q\ra \infty$ is 
$(\a_\mathrm{min}-3)\,q < 0$. On~the other hand, the~exponent vanishes for $q=0$ and $q=1$. 
We deduce that the sequence of integral statistical moments has a reasonable behavior: 
as regards the $r$-dependence, the~exponent in Equation~(\ref{tq}) is negative, 
 thus the moments decrease with $r$; and as regards the asymptotic $q$-dependence, it is 
such that $P(\rho_r)$ is determined by the integral moments, 
according to the standard criteria for a probability distribution function 
to be determined by its integral moments
(p.~20, \cite{Shohat}). 
This property of determinacy cannot be generalized to arbitrary random multiplicative cascades. 

A remark is in order here. People familiar with the theory of critical phenomena 
may find that the description of a 
statistically homogeneous, isotropic and scale invariant mass distribution 
in this section is strange or restrictive, because~the same symmetries are generally 
present in the critical phenomena of statistical physics,
yet the treatment is quite different and the concept of 
multifractal is not necessary. The~crucial point is that we are demanding here 
the scale symmetry of the {\em full} mass distribution and not just the scale symmetry 
of the mass {\em fluctuations} about a homogeneous and isotropic mass distribution. 
The importance of this distinction in cosmology is discussed in~\cite{I0}.

\section{Gravity and Scale~Symmetry}
\label{gravity}

Let us consider the universe on large scales as a system of bodies in gravitational 
interaction and with a moderate range of velocities (the bodies may be galaxies, but~this 
is not important). From~such \mbox{a simple} model, 
we can draw an interesting conclusion in Newtonian gravity (\S 9, \cite{Mandel}):
the accumulated mass $m(r)$ inside a sphere of radius $r$ centered on one body is 
proportional to $r$, that is to say, the~system of bodies approaches, in~a 
range of scales, a~monofractal of dimension one [see Equation~(\ref{mr})]. 
\mbox{The argument} is also very simple: the average velocity of the bodies on the surface 
of the sphere of radius $r$ is about $[G m(r)/r]^{1/2}$, and~this velocity must be 
almost constant. Of~course, this argument ignores that any realization of a 
monofractal of dimension one has to be anisotropic. 
The relation $m(r)\propto r$ is anyway interesting and actually applies 
in various situations and goes beyond 
Newtonian gravity. For~example, the~mass of a black hole or radius $r$ is proportional 
to $r$ (the crucial velocity is then the velocity of light). 
{A more interesting example is the singular solution of the relativistic Oppenheimer--Volkoff equation for isothermal hydrostatic equilibrium, which yields $m(r)\propto r$
(p.~320, \cite{Weinberg_g}).}

The quantity $Gm(r)/r$ represents minus the value of the gravitational potential on the surface of the sphere, assuming isotropy (if there is anisotropy, then consider 
the average over the surface of the sphere). 
In a multifractal distribution of mass in the universe, we may think that the condition 
that the gravitational potential is almost constant everywhere is too restrictive, but~
we should require that it be bounded. Under~this relaxed condition, $\a=1$ becomes 
a lower bound to the local dimension and this bound seems to agree with 
the analysis of data from cosmological simulations and observations of the galaxy distribution~\cite{AinA} {(see the $\a=1$ lower bound in Figure~\ref{Bolshoi})}.
If we further relax the condition of a bounded gravitational potential, the~
dimension one appears again: the set of points for which the potential diverges 
must have a (Hausdorff) dimension smaller than or equal to one (\S 18.2, \cite{Falcon}). 
In~the cosmic web produced by the adhesion model, the~set of filaments and 
point-like singularities is the set of a diverging potential and has precisely 
dimension equal to one. However, the~analysis of data in~\cite{AinA} favors 
a mass distribution in which the potential is either finite everywhere or its set of 
singularities is much more reduced (the precise meaning of this is explained below).

The above preliminary considerations are interesting but have limited predictive power; 
unless one takes too seriously the model of bodies with a restricted range of velocities and concludes that the mass distribution must adjust to a monofractal that is irregular but 
one-dimensional. \mbox{However, Mandelbrot (\S 9, \cite{Mandel})}, who put  forward the argument, 
also asked why the observational exponent of growth of $m(r)$ is larger than one (he quoted  
the value $1.23$). In~retrospect, 
we can affirm that Mandelbrot's 
assumption of a monofractal distribution was the main stumbling block for 
a better understanding of the issue. Indeed, in~a multifractal, $m(r)$ 
has different exponents of growth at different points, and~the analysis of the 
correlation function of galaxies, from~which the value $1.23$ was obtained, only 
yields a sort of average of them. There are reasons to expect multifractality, 
\mbox{apart from} that it realizes the most general form of scale symmetry.  
One reason for multifractality lies in the predictions of the adhesion model, 
which is a reasonably  good model for the early 
formation of structure, including matter sheets and the voids they leave in between, 
as well as filaments and point-like~singularities. 

{Nevertheless, the~filaments and point-like singularities are more complex objects than 
the ones predicted by the adhesion model~\cite{AinA}.
However, this pertains
to Newtonian gravity, while
the formation of matter filaments ({\em thread-like singularities}) and point-like
singularities is feasible in General Relativity  
(Box 32.3, \cite{QG}). 
One may wonder why not consider the problem of 
structure formation in General Relativity from the 
start. General Relativity has no intrinsic length scale, \mbox{similar to} Newtonian gravity, and~should 
also give rise to a fractal structure. 
Actually, General Relativity has indeed been employed in the study of 
large-scale structure formation. 
For example, in~the old cosmology models with 
locally inhomogeneous but globally homogeneous
{spacetime}s that everywhere satisfy Einstein's field equations~\cite{Vittie,Einstein,Einstein+} 
(modernly called ``Swiss cheese'' models). However, these models are far from realistic, 
because they disregard that the initial conditions of structure formation cannot lead to such 
\mbox{a structure}. In~fact, large-scale structure formation can be studied within 
simple Newtonian gravity~\mbox{\cite{Pee,Padma,Peebles}}, that is to say, with~the exception of 
zones with strong gravitational fields, \mbox{where zero-size} singularities can form. However, such 
zones have a very small size, in~cosmic terms. 
}

{The adhesion model is a very simplified model of the action of gravity in 
structure formation, although~it gives a rough idea of the type of early structures.} 
Naturally, what we need for definite and accurate predictions of  
structure formation are dynamic models that are less simplified than the adhesion model.
One can take the full set of cosmological equations of motion in the Newtonian limit
~\cite{Pee,Padma,Peebles}.    
They are applicable on scales small compared to the Hubble length and away from strong gravitational fields, but~they are {\em nonlinear} and quite intractable. 
In fact, these equation 
bring in the classic and hard problem of 
{\em fluid turbulence}, with~the added complication of the gravitational interaction~\cite{AinA}. In~fact, methods of the theory of fluid turbulence can be applied in  
the theory of structure formation, and~it happens that the peculiarities of the 
gravitational interaction can be useful as~constraints. 

For example, let us consider the {\em stable clustering} hypothesis, 
proposed by Peebles and collaborators for the strong clustering regime~\cite{Pee}.   
It says that the average relative velocity of pairs of bodies vanishes. 
This hypothesis arose in the search for simplifying hypothesis to solve a statistical 
formulation of the cosmological equations, one of which was a scaling ansatz. 
Actually, the~stable clustering hypothesis can be considered in its own right and is 
equivalent to the constancy in time of the 
average conditional density, namely   the average density at
distance $r$ from an occupied point. In~a monofractal, it is constant and equals 
the derivative of $m(r)$ divided by the area of a spherical shell of radius $r$ 
\cite{Cole-Pietro,Sylos-PR}. However, just the constancy of the average conditional density 
does not necessarily imply scale symmetry. Nevertheless, the~fact that the average 
conditional density is singular and indeed a power law of $r$ can be argued on 
general grounds~\cite{AinA}.

While the preceding arguments somewhat justify the scale symmetry of the mass distribution, 
they may not be cogent enough and do not predict a definite type of mass distribution.
\mbox{Of course}, the~problem is that we have hardly considered the consequences of the 
equations of motion. The~statistical formulation of these equations is very complicated 
but it can be simplified in the {\em mean field limit}, valid for a large number of 
interacting bodies. In~this limit, the~one-particle distribution function suffices, and~it 
fulfills the {\em collisionless Boltzmann equation} (or Vlasov equation) 
(\S 1.5, \cite{Padma})~\cite{Bavaud}. 
\mbox{Although this} equation is time-reversible, it embodies the nonlinear and chaotic nature 
of the gravitational dynamic and leads to {\em strong mixing}: the flow of matter 
becomes {\em multistreaming} on ever decreasing scales and eventually a state 
of dynamic equilibrium arises, in~a coarse-grained approximation~\cite{GZ}. 
These dynamic equilibrium states fulfill the virial theorem, as~is to be expected and 
is explicitly proved by the {\em Layzer--Irvine equation} (pp.~506--508, \cite{Peebles}). 
Naturally, the~formation of dynamic equilibrium states can be considered as a 
concrete form of the stable clustering~hypothesis.

An equilibrium state that fulfills the virial theorem is not necessarily a 
state of {\em thermodynamic equilibrium}. However, stationary solutions of the 
collisionless Boltzmann equation often mimic the properties of 
the states of thermodynamic equilibrium. As~a case in point, let us take the 
state of equilibrium of an isolated singularity with spherical symmetry, 
obtained by Gurevich and Zybin~\cite{GZ}.
Its density is given by
\begin{equation}
\rho(r) \propto r^{-2}\,[\ln(1/r)]^{-1/3}.
\end{equation}

This density profile can be compared with the density profile $\rho(r) \propto r^{-2}$ 
of the {\em singular isothermal sphere} (which is the asymptotic form of the density profile 
of any isothermal sphere for large $r$) \cite{Padma,Bavaud}. The~difference between 
the two profiles is very small. 
Let us remark that the density profile $\rho(r) \propto r^{-2}$ corresponds to 
the mass--radius relation $m(r)\propto r$ but is unrelated to any fractal property, 
because it applies to a smooth distribution with just one singularity at $r=0$, 
unlike a fractal, in~which every mass point is singular. A~smooth 
distribution of matter, especially    dark matter, with~possibly \mbox{a singularity} at its center, 
is called in cosmology a {\em halo}, and~it has been proposed that 
the large scale structure should be described 
in terms of halos ({\em halo models}) \cite{CooSh}. 
Certain distributions of halos can be considered as coarse-grained multifractals~\cite{fhalos} and, in~fact, halo models and fractal models have much in common~\cite{AinA}.

Given the success of the assumption of thermodynamic equilibrium, 
we take the corresponding 
description of gravitationally bound states 
as a reasonable approach, although~the temperature might have, in~the end, 
a different meaning than it does in the thermodynamics of systems of particles that 
interact through a short range potential. The~thermodynamic approach leads to the 
Poisson--Boltzmann--Emden equation:
\begin{equation}
\Delta \phi = 4\pi G A \exp[-\phi/T],
\label{PBE}
\end{equation} 
where $\phi$ is the gravitational potential, $A$ is a normalizing constant, and~
{$T$ is the temperature in units of energy per unit mass 
(equal to one third of the mean square velocity of bodies)}.
Several derivations of the equation appear in~\cite{Bavaud} (see also (\S 1.5, \cite{Padma}),
\cite{deVega}). 
Probably, the~simplest way of understanding this equation is to realize that it is 
simply the Poisson equation with a source $\rho = A \exp[-\phi/T]$ given by hydrostatic 
equilibrium in the gravitational field defined by $\phi$ itself 
{(for the theory of thermodynamic equilibrium in an external field, 
see (\S 38, \cite{LL})).} 

The solutions of Equation~(\ref{PBE}) depend on the boundary conditions, of~course. 
Simple solutions are obtained by imposing rotational symmetry, namely   the already 
mentioned {\em isothermal spheres} \cite{Padma,Bavaud}. 
Actually, each solution belongs to a family of solutions related by the scale covariance~\cite{deVega} 
\begin{equation}
{\phi_\l(\bm{x}) = \phi(\l \bm{x}) - T\, \log \l^2.}
\end{equation} 

Thus, the~scale-symmetric singular isothermal sphere can be considered as the limit of 
regular isothermal spheres in the same family of solutions. 
({{Let us recall that a singular isothermal sphere with $\rho(r) \propto r^{-2}$  is also a solution in General Relativity, although~the equation for thermodynamic equilibrium 
is more complicated than Equation~(\ref{PBE}) (p.~320, \cite{Weinberg_g}).}})
One can also obtain the 
symmetric solutions corresponding to lower dimension; that is to say, if~we 
take Equation~(\ref{PBE}) in $\mathbb{R}^3$, one obtains the one- or two-dimensional 
solutions by making $\phi$ depend only on one or two variables. These solutions represent 
a two-dimensional sheet or a one-dimensional filament, the~early structures in the adhesion model. More complex solutions of Equation~(\ref{PBE}) 
can be obtained as functions that asymptotically are combinations of the preceding 
solutions. 
Naturally, there are also totally asymmetric solutions, some of which look like 
those combinations of simple solutions (some numerical solutions appear in~\cite{Plum-W}). 
Notice that the formulation of a partial differential equation 
such as Equation~(\ref{PBE}) assumes some regularity of the function, but~we are 
mainly interested in singular solutions. 
We can interpret Equation~(\ref{PBE}) in a coarse-grained sense, that is to say, as~an equation for the coarse-grained variable $\phi_r$
(or $\rho_r$), and~eventually take the limit $r\ra 0$.
However, we can obtain directly singular solutions if we reformulate Equation~(\ref{PBE}) 
as an integral equation (e.g., in~``the Hammerstein description'') \cite{Bavaud}.

Equation~(\ref{PBE}), in~two dimensions, arises in the differential geometry of surfaces, 
where it is called Liouville's equation. In~this context, it is the equation that rules 
the conformal factor of a metric on some surface {[}of course, the~temperature 
$T$ is not present, and~anyway it is always scalable away in Equation~(\ref{PBE}){]}. 
The higher dimensional generalization of Liouville's equation 
also rules the conformal factor of a metric and is then connected with string theory, 
in which the dilaton is \mbox{a dynamical} field in its own right. 
A deeper connection of these relativistic equations, which appear in theories of 
quantum gravity, with~the Newtonian Equation~(\ref{PBE}) may exist, but~we are only concerned here 
with the interpretation and consequences of Equation~(\ref{PBE}) in the theory 
of structure formation. \mbox{Nevertheless, we can} take advantage of the body of knowledge 
developed from the two-dimensional Liouville's equation and Liouville's field theory, 
in particular, the~theory of random multiplicative cascades~\cite{MFcascades,LiouvilleQG,Mchaos,MandelMF,Harte}.
This theory arose in relation to the {\em lognormal model} of turbulence~\cite{Kolmo62}.
\mbox{The lognormal} model certainly has broader scope and often arises in connection 
with nonlinear~processes. 

Examples of the application of the lognormal probability distribution function in 
astrophysics abound, and~there are often connections between them. 
Hubble found that the galaxy counts in
cells on the sky can be fitted by a lognormal distribution~\cite{Hubble}. 
A more relevant example is Zinnecker's model 
of star formation by hierarchical cloud fragmentation 
(a random multiplicative cascade model) \cite{Zinne}. 
\mbox{In cosmology}, a~lognormal model has been proposed as a plausible approximation to the 
large-scale mass distribution~\cite{Coles-Jones}. 
Even though some of these models are explicitly constructed in terms of 
random multiplicative cascades, the~form in which a scale invariant distribution arises 
is by no means evident. In~fact, a~lognormal 
probability distribution function $P(\rho_r)$ of the coarse-grained density $\rho_r$ does 
not imply by itself any scale symmetry, even if it holds for all $r$. It is just 
the dependence of 
$P(\rho_r)$ on $r$ that gives rise to scale symmetry. To~be precise, 
in general, it is necessary that Equation~(\ref{tq}) holds. In~the lognormal model, 
Equation~(\ref{tq}) is fulfilled for a particular form of $\tau(q)$.

Thus, we are led to the study of random self-similar multiplicative cascades. They are 
\mbox{a generalization} of the non-random self-similar multifractals briefly reviewed in 
the preceding section (Section~\ref{MFgeom}). The~study of random cascades is the task of the next~section.

\section{Random~Cascades}
\label{cascades}

Here, we are not interested in regular solutions of Equation~(\ref{PBE}) but 
in singular solutions, in~particular, in~solutions such that $\rho_r(\bm{x})$ is either 
very small or very large. 
As support for the existence of such solutions, we can argue that 
$\rho = A \exp[-\phi/T]$ is always positive but experiences great fluctuations, being 
close to zero at many points but very large at other points, in~accord with 
the fluctuations of $\phi$.  
We are also interested in solutions that fulfill the principle of 
statistical homogeneity and isotropy of the mass distribution. 
The way to obtain such solutions is by means of the theory of random multiplicative~cascades. 

The construction of deterministic self-similar multifractals introduced in 
Section~\ref{MFgeom} is based on the concept of iterated function system. A~function system 
consists of a set of contracting similarities (with some separation condition). 
In addition, it is defined a set of mass ratios that defines how the initial mass 
is split into the sets that result from the application of the function system. 
There are randomized versions of this construction that are not suitable for our problem. 
For example, one can randomize the distribution of mass on 
the sets that result from the function system, by~changing the order at random without altering the mass ratios~\cite{FalconRAND}. This construction has mathematical interest 
but is not very relevant for us, because~it cannot produce either nonlacunarity
or statistical homogeneity and isotropy.  
We instead focus on the random multiplicative cascades based on 
Obukhov--Kolmogorov's 1962 model 
of turbulence~\cite{Kolmo62,Frisch}. 

Obukhov and Kolmogorov intended to describe spatial fluctuations of the energy 
dissipation rate in turbulence and 
were inspired by Kolmogorov's lognormal law of the size distribution 
in pulverization of mineral ore. 
This process can be essentially described as a multiplicative random cascade. 
In fact, similar models had been introduced earlier in economics, under~the name 
of law of {\em proportionate effect} \cite{Gibrat}. All these models are by nature only 
statistical. 
The connection with fractal geometry is due to Mandelbrot, in~the late 1960s 
(the history is exposed in the reprint book~\cite{MandelMF}). 
He observed that a spatial random cascade process of the type used in turbulence, 
when continued indefinitely,
leads to an energy dissipation generally concentrated on a set of non-integer (fractal) 
Hausdorff dimension.
This property was manifested  in some especially simple cascade models, e.g.,~
the $\b$-model, in~which the concentration set is monofractal~\cite{Frisch}.
However, monofractality implies a singular probability $P(\rho_r)$, because~a 
nontrivial monofractal has to be lacunar. Therefore, $P(\rho_r)$ is 
unlike a lognormal probability distribution 
and is not relevant for models of the cosmic mass distribution. 
Actually, Mandelbrot's own random multiplicative cascades generically 
give rise to multifractal mass distributions~\cite{MandelMF}. These are 
the random cascades of interest~here.

A {\em multiplicative process} is defined as follows. Suppose that we start with 
some positive variable of size $x_0$ that at each step $n$ can grow or shrink, 
according to some positive random variable $W_n$, so that
\begin{equation}
x_n = x_{n-1} W_n\,.
\label{xn}
\end{equation} 

Let us think of $x$ as a mass. 
The proportionate effect consists in that the random growth of $x$ is 
\mbox{a percentage} of its current mass and is independent of its current actual mass. 
One may want $x$ to be as likely to grow as to shrink, on~  average; 
that is to say, its mean value must stay constant. \mbox{Usually, the~
random} variables $W_n$, for~$n=1, \ldots$, are assumed to be independent, 
and even assumed to be independent copies of the same random variable $W$. 
A further condition is that $W$ has moments of any order. In~summary, 
the random variable $W$ is subject to:
\begin{equation}
W \geq 0, \quad \langle W \rangle = 1, \quad \langle W^q \rangle < \infty,\,\forall q >0. 
\label{W}
\end{equation} 

As is easy to notice, the~definition of multiplicative process is analogous 
to the definition of additive process, which gives rise to the central limit theorem 
and the Gaussian distribution. \mbox{However, a~multiplicative} process is very different. 
A simple example shows this: if we take $W$ to be zero or two with equal probability, 
then the product of $N$ instances of $W$ will be zero unless all values of $W_n$ turn out 
to be two, in~which case the product is very large, equal to $2^N$.
The mean square value and the variance of the product are also large and higher moments 
are even larger. In~fact, the~moments of the product of $N$ instances of $W$ 
are the $N$th power of the moments of $W$, given by
\begin{equation}
\log_2 \langle W^q \rangle = q-1.
\end{equation} 

This simple example is actually a particular case of the 
$\b$-model, with~$\b=1/2$ (\S 8.6.3, \cite{Frisch}).

The $\b$-model is somewhat singular, because~it lets $W$ be null with a non-vanishing 
probability. \mbox{We can} easily change this feature by taking 
$W$ to be $2p$ or $2(1-p)$ with equal probability, where~$0<p<1$ ($p=0$ or $p=1$ 
give the case already considered). Now, we have
\begin{equation}
\log_2 \langle W^q \rangle = \log_2 \frac{(2p)^q+(2(1-p))^q}{2} = 
\log_2 [p^q+(1-p)^q] + q-1. 
\end{equation} 

Therefore, the~$q$-moment of the product of $N$ instances of $W$ is given by the expression
\begin{equation}
\langle (\prod_{n=1}^N W_n)^q \rangle = \langle W^q \rangle^N = 2^{N[-\tau(q)+q-1]},
\label{qm}
\end{equation} 
where $\tau(q)$ is given by Equation~(\ref{tqm}), because~
the binomial multiplicative process of Section~\ref{MFgeom} 
is closely connected with this process. The~binomial multifractal is possibly the 
simplest example of nonlacunar self-similar multifractal, but~it is deterministic.  
The connection with the present stochastic process is based on the identification 
of the average over $W$ in the multiplicative process with \mbox{a ``binary''} spatial average 
over the mass distribution in the binomial multifractal. 
Of course, the~binomial multifractal could be randomized by using a  
generalization of the procedure in~\cite{FalconRAND}. 
\mbox{However, this construction} would have undesirable 
features. First, the~binary subdivision procedure lets itself to be noticed in the 
final result, breaking the desired statistical homogeneity (as well as statistical isotropy, 
\mbox{in a three-dimensional} generalization). Second, the~total mass is fixed 
at the initial stage in some given volume and we would rather have that it 
has fluctuations that tend to zero as the volume tends to~infinity.

For the application of multiplicative processes to the description of 
the cosmic mass distribution, we have two options. One is to consider a fixed 
volume and to understand that the mass fluctuation in it grows with time, according 
to the natural generalization of Equation~(\ref{qm}), where $N$ is the number 
of time steps. Another is the standard generalization of the definition 
of multiplicative processes 
for the construction of self-similar multifractals, where the volume 
is reduced in every iteration and the average mass is reduced in the same proportion. 
For example, for~a lattice multifractal in $\mathbb{R}^n$, 
we need to replace in Equation~(\ref{W}) $\langle W \rangle = 1$ with 
$\langle W \rangle = b^{-n}$, where $b \geq 2$ is the number of subintervals 
($b^{-n}$ is the volume reduction factor) \cite{Harte}. 
In general, if~we define
\begin{equation}
\tau(q) = -\log_b\langle W^q \rangle -n, 
\end{equation} 
then, taking into account that $r/r_0=b^{-N}$, after~$N$ steps, and~that 
$\rho_r = m_r/r^3$, we recover Equation~(\ref{tq}) in the case $n=3$.

Let us now see how the lognormal distribution arises from Equation~(\ref{xn}). 
Taking logarithms,
\begin{equation}
\log x_N = \log x_{0} + \sum_{n=1}^N \log W_n\,.
\end{equation}

Assuming that the random variables $\log W_n$ fulfill the standard conditions to 
apply the central limit theorem, this theorem implies that the sum of the variables 
converges to a normal distribution and, therefore, the~quotient $x_N/x_0$ 
converges for large $N$ to a lognormal distribution. 
This argument has been criticized on the basis of the theory of large deviations, 
which just says that the limit distribution is expressed in terms of the 
{\em Cram\'er function}, while the central limit theorem only applies to small 
deviations about its maximum, in~the form of a quadratic approximation 
(\S 8.6, \cite{Frisch}) \cite{MandelMF}. \mbox{This quadratic} approximation is insufficient 
to characterize a full multifractal spectrum and only provides an approximation of it, 
whereas the Cram\'er function is sufficient. 
The insufficiency of the central limit theorem is manifest in that there are 
many different quadratic approximations to the multifractal spectrum. Actually, 
two of them are especially important: the quadratic expansions about 
the two distinguished points of a multifractal, namely   
the point where $\a=f(\a)$, which is the point of 
{\em mass concentration}, or~the point where $f(\a)$ is maximum, which corresponds 
to the {\em support} of the mass distribution (which is the full $\mathbb{R}^3$ in our case, 
 thus $f(\a)=3$) \cite{I4}. Both quadratic approximations 
can be employed but they give different results in general. 
At any rate, a~lognormal multiplicative process can be obtained by just requiring that 
the random variables $W_n$ are lognormal themselves, so that the multifractal spectrum 
has an exact parabolic shape (\S 8.6, \cite{Frisch}) \cite{MandelMF,Harte}. 
However, this lognormal model presents some problems (\S 8.6.5, \cite{Frisch}) \cite{MandelMF}.

A surprising feature of an exactly parabolic multifractal spectrum $f(\a)$ is that 
it prolongs to $f(\a)<0$, so there are negative fractal dimensions. The~general nature of 
this anomaly has been discussed by Mandelbrot~\cite{Mandel_neg}. He says that 
``the negative $f(\a)$ rule the sampling variability'' in \mbox{a random multifractal}. 
Any set of singularities of strength $\a$ 
with $f(\a)<0$ is {\em almost surely} empty, but~Mandelbrot states that those values 
``measure usefully the degree of emptiness of empty sets.'' 
\mbox{My option} in the multifractal analysis of 
experimental and observational cosmic mass distributions is to discard 
the part of the multifractal spectrum with $f(\a)<0$ \cite{I-SDSS}.

One more point of concern is possibly the discrete nature of multiplicative processes, 
which are built on a dyadic or $b$-adic tree from the larger to the
smaller scales 
and produce multifractals that have two main drawbacks: 
they display discrete scale invariance only and 
are not strictly translation invariant (or isotropic in $\mathbb{R}^n$, $n>1$). 
In statistical (or quantum) field theory, the~scale symmetry takes place at 
{\em fixed points} of the renormalization group, which is usually formulated as an 
iteration of a discrete transformation but admits a continuous formulation
~\cite{Wil-Kog}. 
An analogous procedure can be carried out for multiplicative processes. 
A simple example of the continuous formulation of multiplicative processes is 
provided by substituting the random walk equation that produces Brownian motion, namely
\begin{equation}
\frac{dx}{dt} = \xi(t),
\end{equation}
where $\xi(t)$ is a Gaussian process (e.g., white noise), by~
the multiplicative equation
\begin{equation}
\frac{dx}{dt} = \xi(t)\,x.
\end{equation}

The solution of this equation is
\begin{equation}
x(t) = x_0\, \exp\! \int_0^t \xi(s)\,ds.
\end{equation}

It undergoes large fluctuations,  such as  the discrete 
multiplicative processes described above, and~can actually be considered as a continuous 
formulation of the lognormal multiplicative process. 
\mbox{A general} definition of continuous multiplicative processes employs the concept of
log infinitely divisible probability distribution~\cite{Muzy}. 
A particularly useful class of these continuous multiplicative processes is 
given by the log-L\'evy generators (\S 6.3.7, \cite{Harte}).
This is an interesting topic but rather technical, so we do not dwell on~it. 

In summary, we have a construction of multifractals that are statistically 
homogeneous and isotropic (in $\mathbb{R}^n$, $n>1$) and have 
continuous scale invariance. Furthermore, they satisfy Equation~(\ref{tq}), for~some 
function $\tau(q)$. 
In fact, we can consider the multifractal as a scale symmetric mass distribution 
that is obtained 
as the continuous parameter $t \ra \infty$. In~this sense, it is analogous to 
a statistical system at a critical point defined in terms of a fixed point of 
the renormalization group. 
However, let us insist that a multifractal is {\em fully}
scale symmetric, whereas a critical statistical system is only required to have
scale symmetric {\em fluctuations}.
Nevertheless, the~mass fluctuations in a continuous scale multifractal also 
tend to zero as $t \ra -\infty$ (at very large scale). 
Therefore, we may consider the transition from the 
fully homogeneous and isotropic cosmic mass distribution on very large scales to 
the multifractal distribution on smaller yet large scales as a 
{\em crossover} analogous to the ones that take place in the theory of 
critical phenomena in statistical physics. 

We must also notice that the scale parameter $t$, analogous to a 
renormalization group parameter, can be replaced by a real scale, such as the 
coarse-graining scale $r$ of Equation~(\ref{tq}). Naturally, $r/r_0 = e^{-t}$, as~
a generalization of $r/r_0=b^{-N}$ (the continuous limit can be thought of as the limit 
where $b\ra 1$ and $N \ra \infty$, with~$N \ln b$ finite). The~use of 
coarse-grained quantities is necessary to connect with the 
partial differential equation (Equation \eqref{PBE}). 
This connection is explained in detail in the collective work 
of \mbox{a group} or researchers in probability theory~\cite{MFcascades,LiouvilleQG,Mchaos,MFcascades0,log-correl}. 
Here, we just need to notice the relation of \mbox{a log-correlated} field, with~\begin{equation}
\langle \phi(x) \phi(y) \rangle = -\log\left| x-y \right|
\end{equation}
for small $\left| x-y \right|$, to~the multifractal cascades. 
This relation comes from the mass density being
\begin{equation}
\rho = A \exp[-\phi/T]
\end{equation}
in Equation~(\ref{PBE}), so that the correlations of the density field are power laws and, 
in particular, Equation~(\ref{tq}) holds. Of~course, this relation cannot specify 
the multifractal properties, given by the function $\tau(q)$. The~case of a 
lognormal distribution  was  studied by Coles and Jones~\cite{Coles-Jones}, 
\mbox{without making} connection with Equation~(\ref{PBE}). 
The lognormal model 
has a quadratic function $\tau(q)$, \mbox{whose Legendre} transform gives the 
parabolic multifractal spectrum $f(\a)$ (\S 6.3.16, \cite{Harte}).


\section{Experimental and Observational~Results}
\label{data}

Fortunately, nowadays we have good knowledge of the cosmic mass distribution, 
based on the considerable amount of data from cosmological simulations and 
from observations of the galaxy distribution. These data are suitable for 
various statistical analyses and, in~particular, for~multifractal analyses. 
Naturally, the~highest quality data come from cosmological $N$-body simulations. 
\mbox{There has} been a steady increase in the number of bodies $N$ that computers are 
capable of handling, and~state-of-the-art simulations handle billions of particles, and~
  thus afford excellent mass resolution. Moreover, $N$-body simulations have 
a numerical experimentation capability, because~one can tune several aspects 
of the cosmic evolution, such as the initial conditions, the~content of baryons in 
relation to dark matter, etc.

\begin{figure}
\centering
\includegraphics[width=7.4cm]{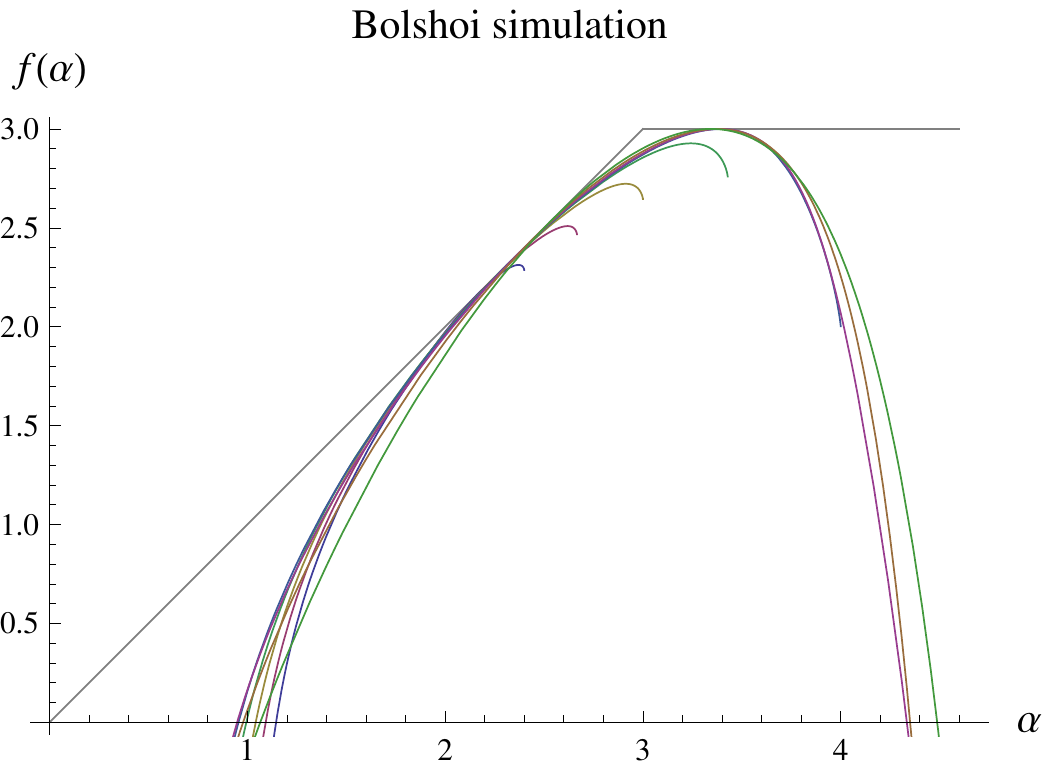}\hspace{5mm}
\includegraphics[width=7.4cm]{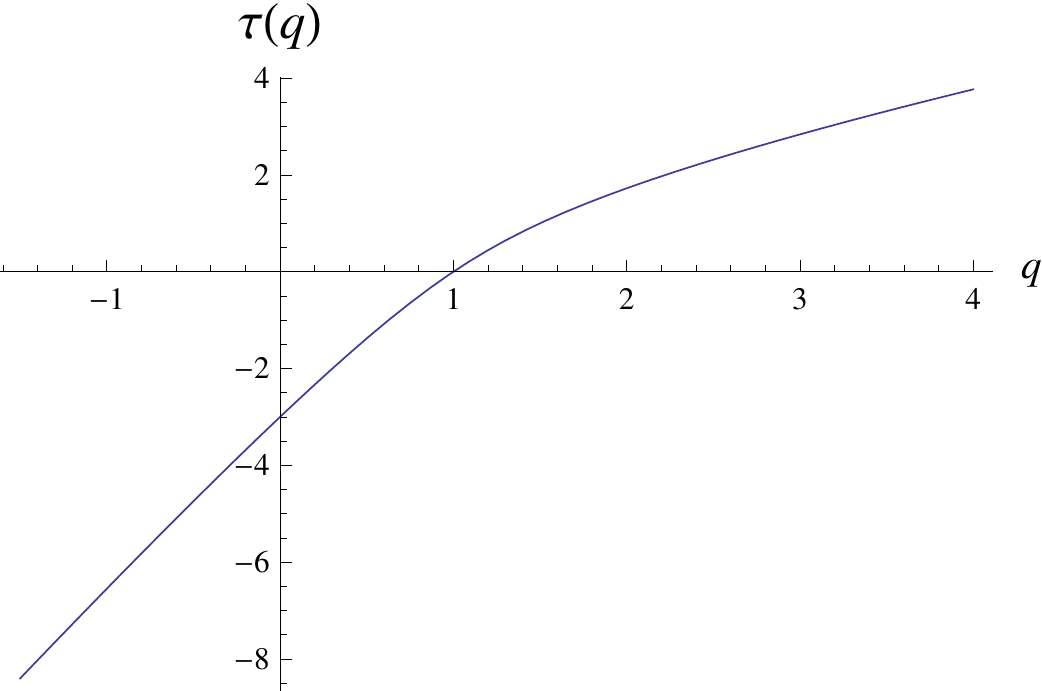}
\caption{(\textbf{Left}) The multifractal spectra of the dark matter distribution in the Bolshoi simulation (coarse-graining lengths $l=3.91, 1.95, 0.98, 0.49, 0.24, 0.122, 0.061, 0.031
\;\mathrm{Mpc}/h$).
(\textbf{Right}) The function $\tau(q)$ (calculated at $l=0.98\;\mathrm{Mpc}/h$), showing that 
$\tau(0)=-3$, corresponding to nonlacunarity. \mbox{This property} is also patent in 
the left graph, because~it shows that $\mathrm{max}\, f(\a) =3$ (for sufficiently 
large~$l$).}
\label{Bolshoi}
\end{figure}

A number of multifractal analyses of the cosmic mass distribution were made years ago~\mbox{\cite{Cole-Pietro,Borga,Valda,Colom,Yepes}}. However, the~quality of the data then was not 
sufficient to obtain reliable results.
\mbox{I have} carried out multifractal analyses 
of the mass distribution in recent several $N$-body simulations, which have consistently 
yielded the same shape of the multifractal spectrum. Probably, the~most interesting 
results to quote are the most recent ones, from~the {\em Bolshoi} simulation~\cite{AinA,Bol}. 
This simulation has very good mass resolution of the cosmic structure:   
it contains $N=2048^3$ particles in a volume of $(250\,\mathrm{Mpc})^3/h^3$ 
[$h$ is 
the Hubble constant normalized to $100$ km$/$(s Mpc)];  
so that each particle represents a mass of $1.35 \cdot 10^8\,h^{-1} M_\odot$, 
which is the mass of a small galaxy. 
Therefore, it is possible to establish convergence of several coarse-grained spectra 
to a limit function, as~the coarse-graining length is shrunk to its minimum 
value available in the particle distribution. Actually, the~shape of 
the multifractal spectrum is stable along a considerable range of scales, 
proving the self-similarity of the mass distribution (Figure~\ref{Bolshoi}). 
It is to be noticed that the 
Bolshoi simulation only describes dark matter particles. However, 
the {\em Mare-Nostrum} simulation describes both dark matter and baryon gas particles 
and gives essentially the same results~\cite{MN}. 

Of course, the~most realistic results about the cosmic mass distribution must come 
from observations of the real universe. Even though there are increasingly better 
observations of the dark matter, the~main statistical analyses of the overall 
mass distribution come from observations of the galaxy distribution. 
I have carried out a multifractal analysis of 
the distribution of stellar mass employing the rich Sloan Digital Sky Survey 
(data release 7) \cite{I-SDSS}. The~stellar mass distribution is \mbox{a proxy} of 
the full baryonic matter distribution and is simply obtained from 
the distribution of galaxy positions by taking into account the stellar masses of galaxies, 
which are available for the Sloan Digital Sky Survey (data release 7).

We can assert a good concordance between 
the multifractal geometry of the cosmic structure in cosmological $N$-body simulations 
and galaxy surveys, to~the extent that the available data allow us to test it, 
that is to say, 
in the important part of the multifractal spectrum $f(\a)$ up to its maximum 
(the part such that $q>0$) \cite{AinA,I-SDSS}. The~other part, such that $q<0$, has 
$\a>3$ and would give information about voids in the stellar mass distribution, 
but the resolution of the SDSS data is insufficient in this range. 
The common features of the multifractal spectrum found in Ref.~\cite{I-SDSS} 
and visible in Figure~\ref{Bolshoi} are:
(i) a minimum singularity strength $\a_\mathrm{min} = 1$; 
(ii) a ``supercluster set'' of dimension $\a=f(\a)\simeq 2.5$ where the mass concentrates; 
and (iii) $\mathrm{max}\, f(\a) =3$, giving a non-lacunar structure 
(without totally empty voids).
As regards Point (i), it is to be remarked that $\a_\mathrm{min} = 1$, 
with $f(\a_\mathrm{min}) =0$, \mbox{corresponds to} the edge of diverging gravitational potential. 
However, the~multifractal spectrum $f(\a)$ prolongs to $f(\a)<0$, giving rise to 
stronger singularities, which have null probability of appearing in the limit of 
vanishing coarse-graining length, $l\ra 0$ (a set with negative dimension is 
{\em almost surely} empty). Nevertheless, 
these strong singularities do appear in any coarse-grained mass distribution 
and correspond to negative peaks of the gravitational potential $\phi$, which 
must not be divergent in the $l	\ra 0$ limit. 
Thus, we seem to have a mass distribution in which the potential can become  large 
(in absolute value) but is finite everywhere (recall the options brought up in Section~\ref{gravity}). 
 
Given that we now know the multifractal spectrum of the cosmic mass distribution with 
reasonable accuracy, we can look for the type of random multiplicative cascade that 
produces such spectrum. This is an appealing task that is left for the~future.

\section{Discussion}

We   show that there is a considerable range of scales in the universe in which 
scale symmetry is effectively realized, that is to say, the~mass distribution is a 
self-similar multifractal, with~identical appearance and properties at any scale. 
This symmetry is a consequence of the absence of any intrinsic length scale in 
Newtonian gravitation, which is the theory that rules the mass distribution 
on scales beyond the size of galaxies but small compared to the Hubble length. 
Indeed, it is found in the analysis of cosmological $N$-body simulations and in 
the analysis of the stellar mass distribution (with less precision) 
that the self-similarity extends from a fraction of Megaparsec to several Megaparsecs. 
\mbox{On larger} scales, the~multifractal mass distribution shows signs of undergoing 
the transition to the expected homogeneity 
of the Friedmann--Lemaitre--Robertson--Walker relativistic model of the universe, 
\mbox{in accord} with the standard cosmological~principle. 

We   describe  models that enforce scale symmetry in combination with the other 
relevant symmetries, namely  the translational and rotational symmetries that impose 
homogeneity and isotropy and that must be understood in a 
statistical sense, related to Mandelbrot's conditional cosmological principle. 
We    show that those models are given by the theory of 
continuous random multiplicative cascades. We   show how 
random multiplicative cascades can be constructed, made continuous, and~produce the 
type of multifractal mass distribution that we~need. 

Of course, the~use of continuous random multiplicative cascades could be regarded 
as somewhat {  ad~hoc}, as~they seem unrelated to the gravitational physics. 
However, we have explained the close connection of those models with the 
partial differential equation that arises from an approximate model of 
gravitational physics, namely   the Poisson--Boltzmann--Emden equation {that follows from} 
the assumption of thermodynamic equilibrium. While this assumption may not be 
fully realized in the cosmological evolution of structure formation, it is a 
reasonable approach to the states of virial equilibrium, say, to~
the regime of strong and stable~clustering. 

We   also mention  that the early stage of structure formation is approximately 
described by the adhesion model, which predicts a self-similar cosmic web somewhat 
different from the result of \mbox{a continuous} random multiplicative cascade related 
to the Poisson--Boltzmann--Emden equation, as~can be perceived in 
Figure~\ref{devil}. 
It seems that both types of structures should be combined, \mbox{being the} web morphology
appropriate on the larger scales and the full self-similar multifractal structure,  
related to the Poisson--Boltzmann--Emden equation, appropriate on smaller scales. 
This combination should take into account that the matter
sheets and the corresponding voids present no problem in Newtonian gravity, whereas the 
matter filaments and point-like singularities do and they should be replaced 
by weaker singularities, of~power law type, precisely such as the ones 
that are found as simple solutions of the Poisson--Boltzmann--Emden equation; 
namely, radial or axial singular isothermal distributions. 
\mbox{The combined} structure achieves a mass distribution without singularities of 
the gravitational~potential. 

Since the Poisson--Boltzmann--Emden equation describes halo-like structures 
(Section~\ref{gravity}), we can expect that a coarse-grained formulation of 
the above proposed combination of singular solutions of the 
Poisson--Boltzmann--Emden equation with a larger-scale web structure would be 
equivalent to the fractal distribution of halos that can be deduced 
from a coarse-grained multifractal~\cite{fhalos,I4}. \mbox{Moreover, such a
distribution} of halos should lie in web sheets or filaments, as~
is expected~\cite{CooSh,Bol}.


Finally, we can just mention here that the formalism described in this paper can 
possibly be applied in a very different range of scales and with a different 
theory of gravity, namely  the very small scales that constitute the realm of 
quantum gravity. The~potential of scale symmetry and, \mbox{furthermore, of~
conformal} symmetry in theories of quantum gravity is well established and, in~fact, 
string theory is essentially based on the conformal symmetry. 
Anyway, this very interesting connection lies beyond the scope of the present~work.

\vspace{6pt} 


\funding{This research received no external funding.}

\conflictsofinterest{The author declares that there is no conflict of interest regarding the publication of this~paper.}

\reftitle{References}

\end{document}